\documentclass[10pt,fleqn]{article}
\usepackage{amsmath}
\usepackage{amssymb}
\usepackage{mathtools}
\usepackage{geometry}
\usepackage{newpxtext}
\usepackage{newpxmath}
\usepackage{dsfont} 
\usepackage{caption} 
\usepackage{subcaption}
\usepackage{mathrsfs}	
\usepackage{cite} 
\usepackage{enumitem}
\usepackage{array}
\usepackage{booktabs}
\usepackage{xurl}
\usepackage{nicefrac}
\usepackage[linkcolor=blue,urlcolor=blue,citecolor=magenta,colorlinks=true]{hyperref}
\usepackage[noabbrev]{cleveref}
\usepackage[linkcolor=blue,urlcolor=blue,citecolor=magenta,colorlinks=true]{hyperref}
\usepackage{graphicx}
\usepackage{scalerel}
\usepackage[nolist]{acronym}
\usepackage{braket}
\usepackage{mleftright}\mleftright
\usepackage{booktabs}
\usepackage{accents}
\usepackage{xfrac}
\usepackage{xspace}
\usepackage{enumitem}
\usepackage[normalem]{ulem}
\usepackage{soul}
\usepackage{microtype}
\makeatletter
\g@addto@macro\bfseries{\boldmath}
\newcommand*{\defeq}{\mathchoice{\mathrel{\rlap{%
\raisebox{0.24ex}{$\m@th\cdot$}}%
\raisebox{-0.24ex}{$\m@th\cdot$}}%
=}{\mathrel{\rlap{%
\raisebox{0.24ex}{$\m@th\cdot$}}%
\raisebox{-0.24ex}{$\m@th\cdot$}}%
=}{\mathrel{\rlap{%
\raisebox{0.08ex}{\small$\m@th\cdot$}}%
\raisebox{-0.28ex}{\small$\m@th\cdot$}}%
=}{\mathrel{\rlap{%
\raisebox{0.08ex}{\tiny$\m@th\cdot$}}%
\raisebox{-0.28ex}{\tiny$\m@th\cdot$}}%
=}}
\newcommand*{\eqdef}{\mathchoice{=\mathrel{\rlap{%
\raisebox{0.24ex}{$\m@th\cdot$}}%
\raisebox{-0.24ex}{$\m@th\cdot$}}}{%
=\mathrel{\rlap{%
\raisebox{0.24ex}{$\m@th\cdot$}}%
\raisebox{-0.24ex}{$\m@th\cdot$}}}{%
=\mathrel{\rlap{%
\raisebox{0.08ex}{\small$\m@th\cdot$}}%
\raisebox{-0.28ex}{\small$\m@th\cdot$}}}{%
=\mathrel{\rlap{%
\raisebox{0.08ex}{\tiny$\m@th\cdot$}}%
\raisebox{-0.28ex}{\tiny$\m@th\cdot$}}}%
}
\newcommand*{\transpose}{
    {\mathpalette\@transpose{}}%
    }
\newcommand*{\@transpose}[2]{%
    \raisebox{\depth}{$\m@th#1\intercal$}%
}
\makeatother
\newlist{requirements}{enumerate}{1}
\setlist[requirements]{label=(\Roman*),ref=(\Roman*),leftmargin=*, widest=abcd}
\crefname{requirementsi}{requirement}{requirements}
\Crefname{requirementsi}{Requirement}{Requirements}

\allowdisplaybreaks
\newcommand{\dd}{\mathop{}\!\mathrm{d}}
\newcommand{\ii}{\mathop{}\!\mathrm{i}\!\mathop{}}
\newcommand{\ee}{\mathrm{e}}
\newcommand\simnsam{\mathrel{\ooalign{$\simeq$\cr
  \hidewidth\raise-.333ex\hbox{\rotatebox{45}{$\shortmid$}}\hidewidth\cr}}}
\DeclareMathOperator{\re}{Re}
\DeclareMathOperator{\im}{Im}

\DeclareMathOperator{\diag}{diag}

\newcommand{\ChargeC}{\ensuremath{\mathcal{C}}}

\newcommand{\ParityP}{\ensuremath{\mathcal{P}}}
\newcommand{\CP}{\ensuremath{\ChargeC\ParityP}\xspace}

\newcommand{\rep}[1]{\ensuremath{{\boldsymbol{#1}}}}

\newcommand{\Z}[1]{\ensuremath{\mathds{Z}_{#1}}}
\newcommand{\deltaCP}{\ensuremath{\delta}}

\newcommand{\RequirementSymbol}[2][]{\scalerel*{\ensuremath{\vcenter{\hbox{\includegraphics{r#2.pdf}}}}}{\ensuremath{M^1_2}}}
\newcommand{\Requirement}[2][]{\ifcase#2\relax\or 
\hyperref[item:modular_invariance]{\RequirementSymbol{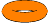}}\xspace
\or 
\hyperref[item:depend_on_tau_only]{\RequirementSymbol{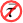}}\xspace
\or 
\hyperref[item:finite]{\RequirementSymbol{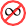}}\xspace
\fi}%
\begin{document}
\title{%
\begin{flushright}
\mbox{\small\texttt{UCI-TR-2023-11}}
\end{flushright}
\par\bigskip\bigskip
{\LARGE\bfseries
Modular invariant holomorphic observables}}
\date{\small January 2024}
\author{\begin{minipage}{0.85\linewidth}\normalsize\centering
Mu-Chun Chen$^{a}$\footnote{E-mail: \texttt{muchunc@uci.edu}},  \
Xiang-Gan Liu$^{a}$\footnote{E-mail: \texttt{xianggal@uci.edu}},  \
Xue-Qi Li$^{a}$\footnote{E-mail: \texttt{xueqi.li@uci.edu}},  \
Omar Medina$^{b}$\stepcounter{footnote}\footnote{E-mail: \texttt{omar.medina@ific.uv.es}}, \
Michael Ratz$^{a}$\stepcounter{footnote}\footnote{E-mail: \texttt{mratz@uci.edu} \
}
\end{minipage}
\\*[20pt]
\begin{minipage}{0.7\linewidth}
\begin{center}
{\itshape\small$^a$ Department of Physics and Astronomy, University of California, Irvine, CA~92697-4575, USA}\\[5mm]
{\itshape\small$^b$ Instituto de Física Corpuscular (IFIC), Universidad de Valencia-CSIC,\\
Paterna (Valencia) E-46980, Spain}
\end{center}
\end{minipage}
}

\maketitle
\thispagestyle{empty}

\begin{abstract}
In modular invariant models of flavor, observables must be modular invariant.
The observables discussed so far in the literature are functions of the modulus $\tau$ and its conjugate, $\bar\tau$.
We point out that certain combinations of observables depend only on $\tau$, i.e.\ are meromorphic, and in some cases even holomorphic functions of $\tau$.
These functions, which we dub ``invariants'' in this Letter, are highly constrained, renormalization group invariant, and allow us to derive many of the models' features without the need for extensive parameter scans. 
We illustrate the robustness of these invariants in two existing models in the literature based on modular symmetries, $\Gamma_{3}$ and $\Gamma_{5}$. 
We find that, in some cases, the invariants give rise to robust relations among physical observables that are independent of $\tau$. 
Furthermore, there are instances where additional symmetries exist among the invariants. 
These symmetries are relevant phenomenologically and may provide a dynamical way to realize symmetries of mass matrices.
\end{abstract}
\clearpage

\section{Introduction}
\label{sec:Introduction}

The \ac{SM} contains, including neutrinos, almost 30 continuous parameters. 
Most of these parameters reside in the flavor sector. 
Recently, the modular invariant approach to flavor has emerged as a promising way of deriving the flavor parameters from powerful modular symmetries \cite{Feruglio:2017spp}, with a significantly reduced number of parameters. 
The construction of the flavor models entails the finite modular groups $\Gamma_N$~\cite{Feruglio:2017spp} or $\Gamma_N'$~\cite{Liu:2019khw}.
A number of explicit models has been worked out, such as \cite{Meloni:2023aru,Ding:2019zxk,Novichkov:2018nkm,Ding:2019xna,Liu:2020akv,Novichkov:2020eep,Liu:2020msy,Yao:2020zml,Wang:2020lxk,Ding:2020msi,Li:2021buv,Ding:2023ydy,Arriaga-Osante:2023wnu}, see the recent reviews~\cite{Kobayashi:2023zzc,Ding:2023htn} for further references.

One of the main reasons for the popularity of this scheme is that the couplings of the theory are unique, or at least very constrained. 
In slightly more detail, the couplings are functions of a chiral superfield $\tau$ subject to three requirements: 
\begin{requirements}
 \item[\RequirementSymbol{1}] modular covariance or modular invariance (cf.\ \Cref{sec:ModularInvariance}),\label{item:modular_invariance}  
 \item[\RequirementSymbol{2}] the couplings depend only on the modulus $\tau$ but not on its conjugate, $\bar\tau$, and\label{item:depend_on_tau_only} 
 \item[\RequirementSymbol{3}] the couplings are finite for all values of $\tau$.\label{item:finite}
\end{requirements}
Note that in different communities, different terminology is being used for these requirements.
In mathematics, \Requirement{2} amounts to saying that the coupling is a meromorphic function of $\tau$ whereas in some physics contexts such functions are called holomorphic. 
\Requirement{2} and \Requirement{3} together mean, in mathematician's terminology, that the coupling is a holomorphic function of $\tau$.
In what follows, we will refer to the requirements just by the symbols to avoid confusion.

The important point is that \Requirement{1}, \Requirement{2} and \Requirement{3} together are so restrictive that they almost completely fix the couplings~\cite{ranestad20081,Sebbar:2002,Schultz:2015}.
Up to this point, these requirements have only been discussed at the level of superpotential couplings.
The purpose of this Letter is to point out that one can make similar statements at the level of observables. 
As we shall see, there are observables for which all the requirements, i.e.\ \Requirement{1}, \Requirement{2} and \Requirement{3}, are simultaneously fulfilled. 
Our findings allow us to make very robust predictions. 
There are also observables which fulfill \Requirement{1} and \Requirement{2} but fail to satisfy \Requirement{3}. 
We will comment on how in such cases one can still make general statements on the predictions of the model.

In this Letter, we first review the basic framework of modular flavor symmetries in \Cref{sec:ModularInvariance}. 
In \Cref{sec:Non-holomorphicObservables} we discuss how typical observables are non-holomorphic since they involve the normalization of the fields.
We then introduce modular invariant holomorphic observables in \Cref{sec:ModularInvariantHolomorphicObservables}, and work out some 
basic applications using two example models in \Cref{sec:Feruglio_Model_1,sec:A_5_Model}.
\Cref{sec:Discussion} contains some further discussion, and \Cref{sec:Summary} contains our conclusions.

\section{A short recap of modular flavor symmetries}
\label{sec:ModularInvariance}

The key ingredient of modular flavor symmetries is modular invariance, i.e.\ requirement \Requirement{1}. 
That is, the theory is assumed to be invariant under $\text{SL}(2,\mathds{Z})$ transformations $\gamma$ of the modulus $\tau$,
\begin{equation}\label{eq:modular_transformation_tau}
 \tau\xmapsto{~\gamma~}\frac{a\,\tau+b}{c\,\tau+d}\;,
\end{equation}
where $a,b,c,d\in\mathds{Z}$ and $a\,d-b\,c=1$.
Modular invariance, along with \Requirement{2} from \ac{SUSY}\footnote{\ac{SUSY} may not be necessary for \Requirement{2}, cf.\ \cite{Almumin:2021fbk}. However, so far there is no explicit model illustrating this.} and the additional requirement that the couplings of the theory be finite, i.e.\ \Requirement{3}, leads to a highly predictive scheme, in which the superpotential couplings are almost unique.
That is, the superpotential terms of the models are of the form 
\begin{equation}\label{eq:ModularInvariantW}
 \mathscr{W}\supset  g \,Y_{ijk}(\tau)\,\Phi^i\,\Phi^j\,\Phi^k\;,      
\end{equation}
where $Y_{ijk}(\tau)$ are uniquely determined vector-valued modular forms and the $\Phi^i$ denote some appropriate superfields. 
That is, under \eqref{eq:modular_transformation_tau} 
\begin{equation}
  Y_{ijk}(\tau)\xmapsto{~\gamma~}Y_{ijk}(\gamma\,\tau)=(c\tau+d)^{k_Y}\,\rho_Y(\gamma)\,Y_{ijk}(\tau)\;,
\end{equation}
where $\rho_Y$ is a representation matrix of a finite group. 
As long as $k_Y\ne0$ and/or $\rho_Y(\gamma)\ne\mathds{1}$, $Y_{ijk}(\tau)$ is modular covariant (rather than invariant).
The superfields transform as 
\begin{equation}\label{eq:ModularTransformationOfPhi_i}
  \Phi_i\xmapsto{~\gamma~} (c\tau+d)^{-k_i}\,\rho_i(\gamma)\,\Phi_i\;,  
\end{equation}
where $k_i$ denotes the modular weight of $\Phi_i$, which, as indicated may transform nontrivially under the finite modular group with representation matrix $\rho_i(\gamma)$. 
$g$ denotes a coefficient, which can be chosen arbitrarily in the bottom-up approach.\footnote{See, however, \cite{Petcov:2023fwh} for a proposal for the normalization of the modular forms. It would be interesting to see to which extent this approach replicates the known normalizations in explicit top-down constructions such as \cite{Nilles:2021glx}.}  
However, apart from this freedom, the superpotential terms are uniquely determined by requirements \Requirement{1}, \Requirement{2} and \Requirement{3}. 

Let us briefly recall what \Requirement{3} means. 
It is the requirement that the functions $Y_{ijk}(\tau)$ remains finite throughout the fundamental domain. 
Without this requirement, we could multiply $Y_{ijk}(\tau)$ by arbitrary polynomials of the modular invariant function, or Hauptmodul of $\text{SL}(2,\mathds{Z})$, $j(\tau)$, while still satisfying requirements \Requirement{1} and \Requirement{2}.
However, as $j$ diverges for $\tau\to\ii\infty$, this is inconsistent with \Requirement{3}, and therefore not allowed. Thus, by requiring simultaneously \Requirement{1}, \Requirement{2}, and \Requirement{3}, the couplings are unique up to an undetermined coefficient $g$, and in cases where there are multiple invariant contractions, up to multiple undetermined coefficients $g_{i}$. Examples for the latter case can be found in~\cite{Ding:2019zxk,Liu:2020akv,Li:2021buv}.

In vast literature, the K\"ahler potential of the matter fields is assumed to be of the so-called minimal form,
\begin{equation}\label{eq:MinimalK}
 K_\mathrm{matter}=\sum\limits_i \frac{1}{(-\ii\tau+\ii\bar\tau)^{k_i}}\overline{\Phi_i} \Phi_i\;.      
\end{equation}
Here we set the vector multiplets to zero. 
It is known that the requirements \Requirement{1}, \Requirement{2} and \Requirement{3} do not fix the K\"ahler potential to be of the form~\eqref{eq:MinimalK}, but there are several additional terms which are allowed by the symmetries of the models, thus limiting the predictive power of the models~\cite{Chen:2019ewa}. 
While entirely convincing solutions to this problem have not yet been found, there exist proof-of-principle type fixes  which allow one to sufficiently control the extra terms to make their impact comparable to the current experimental uncertainties in flavor observables \cite{Chen:2021prl}.
In what follows, we will base our discussion on the minimal K\"ahler potential~\eqref{eq:MinimalK}.

Modular invariance, in particular, means that observables are to be modular invariant.
However, this does \emph{not} mean that the $Y_{ijk}(\tau)$ of~\eqref{eq:ModularInvariantW} are modular invariant. 
Rather, as we shall see next, there are several non-holomorphic observables which are modular invariant because of the normalization of the fields, cf.\ \Cref{eq:MinimalK}.

\section{Non-holomorphic observables}
\label{sec:Non-holomorphicObservables}

To illustrate this point with an explicit example, we consider a toy model based on
\begin{subequations}
\begin{align}
 \mathscr{W}&=\frac{\mathcal{M}(\tau)}{2}\,\Phi^2\;,\\
 K&=\frac{1}{(-\ii\tau+\ii\bar\tau)^{k_\Phi}} \overline{\Phi} \Phi
 \; ,
 \label{eq:ToyKahler}
\end{align}  
\end{subequations}
where $\mathcal{M}(\tau)$ is a vector-valued modular form of weight $k_{\mathcal{M}}$.
Apart from the modular weight of the field, $k_\Phi$, we need to specify the modular weight of the superpotential $k_{\mathscr{W}}=k_{\mathcal{M}}+2k_\Phi$. In large parts of the literature, the modular weight of the superpotential is taken to be zero, $k_{\mathscr{W}}=0$, and in this section we adopt this practice. A nonzero $k_{\mathscr{W}}=0$, as required by supergravity, does not change the following discussion qualitatively.
This then fixes the modular weight of $\mathcal{M}(\tau)$ to be 
\begin{equation}
 k_{\mathcal{M}}=-2k_\Phi\;. \label{eq:ToyWeights}
\end{equation}
Thus, under a modular transformation 
\begin{subequations}\label{eq:ToyTransformations}
\begin{align}
 \tau&\mapsto \frac{a\tau+b}{c\tau+d}\eqdef\tau'\;, \label{eq:ToyTauTrans} \\
 \Phi&\mapsto \frac{1}{(c\tau+d)^{k_\Phi}}\Phi\;,\label{eq:ToyPhiTrans}\\
 \mathcal{M}(\tau)&\mapsto \mathcal{M}(\tau')  
 =\frac{1}{(c\tau+d)^{k_\mathcal{M}}}\mathcal{M}(\tau)\;.\label{eq:ToyMTrans}
\end{align}  
\end{subequations}
While $\mathcal{M}(\tau)$ and $\tau$ both transform nontrivially, is straightforward to confirm that 
\begin{equation}
  |\mathcal{M}(\tau)|^2\,(-\ii\tau+\ii\bar\tau)^{-k_{\mathcal{M}}}
\end{equation}
is invariant under~\eqref{eq:ToyTransformations}.
This combination emerges from the scalar potential,
\begin{equation}\label{eq:Vtoy}
 \mathscr{V}=\left(\frac{\partial\overline{\mathscr{W}}}{\partial\overline{\Phi}}\right)\,K^{\overline{\Phi}\Phi}\,\frac{\partial\mathscr{W}}{\partial\Phi}+\dots 
\end{equation}
after rescaling the fields to be canonically normalized.
That is, we have to take into account the inverse of the K\"ahler metric, which we obtain from \Cref{eq:ToyKahler}
\begin{equation}
  K^{\overline{\Phi}\Phi}=(-\ii\tau+\ii\bar\tau)^{k_\Phi}\;,
\end{equation}
both in~\eqref{eq:Vtoy} and when computing the physical mass, 
\begin{equation}
 m_\mathrm{physical}^2=(-\ii\tau+\ii\bar\tau)^{k_\Phi}\,\left.\frac{\partial^2\mathscr{V}}{\partial\Phi\,\partial\overline{\Phi}}\right|_{\Phi=\overline{\Phi}=0} \;. 
\end{equation}
Here, $\Phi$ denotes the scalar component of the superfield $\Phi$.
The resulting physical mass of $\Phi$ is given by
\begin{equation}
  m_\mathrm{physical}=m_\mathrm{physical}(\bar\tau,\tau)= \lvert \mathcal{M}(\tau)\rvert \,(-\ii\tau+\ii\bar\tau)^{k_\Phi}\;.
\label{eq:NonHoloMass}
\end{equation}
As indicated by the notation, the physical mass is \emph{not} a meromorphic (nor holomorphic) function of $\tau$, i.e.\ it does not fulfill \Requirement{2}. 
Of course, the physical mass $m_\mathrm{physical}(\bar\tau,\tau)$ is modular invariant, as it should, i.e.\ satisfies \Requirement{1}.
However, it is modular invariant ``at the expense'' of being non-holomorphic, i.e.\ \Requirement{1} and \Requirement{3} are fulfilled but not \Requirement{2}.

The observable of the model, i.e.\ the mass, fails to satisfy \Requirement{2} because it involves the K\"ahler metric. 
Let us stress that this feature of the toy model is rather generic: in order to compute observables, one typically needs to take into account the K\"ahler metrics, thereby sacrificing \Requirement{2}. As a consequence, the uniqueness discussed in \Cref{sec:ModularInvariance} does not apply to such observables.
In what follows, we will see that there are observables that do not receive $\bar\tau$-dependents contributions, and hence fulfill \Requirement{2}, and are thus highly constrained.

\section{Modular invariant holomorphic observables}
\label{sec:ModularInvariantHolomorphicObservables}

In order to obtain holomorphic observables, we need to remove the nonholomorphic terms coming from the K\"ahler metric.
It turns out that in the lepton sector of the \ac{MSSM} there is a straightforward way to obtain such expressions. 
Consider the superpotential of the lepton sector 
\begin{equation}
  \mathscr{W}_\mathrm{lepton}
  =
  Y_e^{ij}\,L_i\cdot H_d\,E_j
  +
  \frac{1}{2}\kappa_{ij}(\tau)\,L_i\cdot H_u\,L_j\cdot H_u\;.
\end{equation}
Here, $L^i$ and $E^i$ denote the three generations of the superfields of the $\text{SU}(2)_\mathrm{L}$ charged lepton doublets and singlets, and $H_{u/d}$ stand for the \ac{MSSM} Higgs doublets.
$Y_e$ denotes the charged lepton Yukawa couplings, which is not a modular form. $M(\tau)=v_u^2\,\kappa(\tau)$ is the neutrino mass matrix, with  $\kappa(\tau)$ being the effective neutrino mass operator.

In the basis in which $Y_e=\diag(y_e,y_\mu,y_\tau)$, consider 
\begin{equation}\label{eq:Definition_of_I_ij_in_MSSM}
 I_{ij}(\tau)\defeq \frac{M_{ii}(\tau)\,M_{jj}(\tau)}{\bigl(M_{ij}(\tau)\bigr)^2}  
 =
 \frac{\kappa_{ii} (\tau) \,\kappa_{jj} (\tau) }{\bigl(\kappa_{ij} (\tau)\bigr)^2}  
 =
 \frac{m_{ii} (\tau, \bar\tau) \,m_{jj} (\tau, \bar\tau)}{\bigl(m_{ij} (\tau, \bar\tau)\bigr)^2}  
 \;,
\end{equation}
where no summation over $i,j$ is implied.  
Here,  $m_{ij} (\tau, \bar\tau) \defeq \left(-\ii\tau+\ii\bar\tau\right)^{(k_{L_{i}} + k_{L_{j}})/2} \kappa_{ij} (\tau) \,v_u^2$ are the entries of the  neutrino  mass matrix in the canonically normalized basis, with $k_{L_{i}}$ being the modular weight of the lepton doublet $L_{i}$. 
Crucially, \Cref{eq:Definition_of_I_ij_in_MSSM} shows that the ratios of the physical mass matrix entries, $m_{ij}(\tau,\bar{\tau})$, can be expressed entirely as rational functions of \emph{holomorphic} modular functions. 
This is because, while the individual entries $m_{ij}(\tau,\bar{\tau})$ have a structure analogous to \Cref{eq:NonHoloMass}, the $I_{ij}$ are constructed in such a way that the factors containing $\bar\tau$ cancel.
That is, by construction the $I_{ij}$ fulfill \Requirement{2}. 
In what follows, we will discuss to which extent they also fulfill \Requirement{1} and \Requirement{3}.

It has been known for a while that the $I_{ij}$ from \Cref{eq:Definition_of_I_ij_in_MSSM} are \ac{RG} invariant~\cite{Chang:2002yr} (see the discussion in \Cref{sec:RGinvariants}.)   
In the \ac{MSSM} this can be understood from the non-renormalization theorem and the fact that the normalizations of the field cancel, which is also the reason why the $I_{ij}$ from \Cref{eq:Definition_of_I_ij_in_MSSM} are interesting for our present discussion.
As we detail in \Cref{sec:RGinvariants}, the location of zeros and poles are \ac{RG} invariant to all orders even in the absence of \ac{SUSY}. 
In this basis, the neutrino mass matrix is given by 
\begin{equation}\label{eq:m_nu_U_MNS}
m_\nu=U^{*}_\mathrm{PMNS}\,\diag\bigl(m_1,m_2,m_3\bigr)\,U^{\dagger}_\mathrm{PMNS}\;, 
\end{equation}
where the $m_i$ denote the neutrino mass eigenvalues and the PMNS matrix $U_\mathrm{PMNS}$ depends on the leptonic mixing angles ($\theta_{12}, \theta_{13}, \theta_{23}$), the Dirac phase ($\delta$) and the two Majorana phases ($\varphi_{1}, \varphi_{2}$). 
Altogether, there are nine independent physical parameters,
\begin{equation}
\{\theta_{12}, \theta_{13}, \theta_{23}, \deltaCP, \varphi_{1}, \varphi_{2}, m_1, m_2, m_3\}\;.
\label{eq:LeptonMixingParameters}
\end{equation}
From \Cref{eq:m_nu_U_MNS}, the invariants can be computed explicitly, and read, in the PDG basis,\footnote{Note that we have chosen a different notation $\varphi_{i} \defeq -2\eta_i$ for the Majorana phases with respect to the PDG~\cite{ParticleDataGroup:2022pth}.}
\begin{subequations}
\label{eq:RGinvariant}
\begin{align}
 I_{12} & =  
 \frac{a_0\,\left[
 \widetilde{m}_1 \left(\ee^{\ii \deltaCP } c_{23} s_{12}+c_{12} s_{13} s_{23}\right)^2
 +\widetilde{m}_2 \left(\ee^{\ii \deltaCP } c_{12} c_{23}-s_{12} s_{13} s_{23}\right)^2
 +\ee^{2 \ii \deltaCP } m_3 c^2_{13} s^2_{23}
 \right]}{ c^2_{13}\left[
 \widetilde{m}_1 c_{12} \left(\ee^{\ii \deltaCP } c_{23} s_{12}+c_{12} s_{13} s_{23}\right)
 +\widetilde{m}_2 s_{12} \left(s_{12} s_{13} s_{23}-\ee^{\ii \deltaCP }c_{12} c_{23}\right)
 -\ee^{2 \ii \deltaCP } m_3 s_{13} s_{23}
 \right]^2}
 \;,\\
 I_{13} & =  
 \frac{a_0\, \left[
 \widetilde{m}_1 \left(c _{12} c _{23} s _{13}-\ee^{\ii \deltaCP }s _{12} s_{23}\right)^2
 +\widetilde{m}_2 \left(c _{23} s _{12} s _{13}+\ee^{\ii \deltaCP } c_{12} s _{23}\right)^2
 +\ee^{2 \ii \deltaCP } m_3 c^2_{13} c ^2_{23}
 \right]}{c^2_{13}\left[
 \widetilde{m}_1 c _{12} \left(c _{12} c _{23} s _{13}-\ee^{\ii \deltaCP }  s_{12} s _{23}\right)
 +\widetilde{m}_2 s _{12} \left(c _{23} s _{12} s _{13}+\ee^{\ii \deltaCP } c _{12}  s_{23}\right)
 -\ee^{2 \ii \deltaCP } m_3 c _{23} s_{13}
 \right]^2}
 \;,\\
 I_{23} & =  
 \left[\ee^{2 \ii \deltaCP } m_3 c^2_{13} s^2_{23}+\widetilde{m}_1 \left(\ee^{\ii \deltaCP }
 c _{23} s _{12}+c _{12} s _{13} s _{23}\right)^2+\widetilde{m}_2 \left(\ee^{\ii \deltaCP
 } c _{12} c _{23}-s _{12} s _{13} s _{23}\right)^2\right]
 \nonumber\\
 &\qquad  {}\times
 \frac{4 \left[\ee^{2 \ii \deltaCP } m_3 c^2_{13} c^2_{23}+\widetilde{m}_2 
 \left(c_{23} s_{12} s_{13}+\ee^{\ii \deltaCP } c_{12} s _{23}\right)^2+\widetilde{m}_1
 \left(c_{12} c_{23} s_{13}-\ee^{\ii \deltaCP } s_{12} s _{23}\right)^2\right]}{%
 \left[
	\widetilde{m}_1 a_1
	+\widetilde{m}_2 a_2
 	-\ee^{2 \ii \deltaCP} m_3 \sin (2 \theta _{23}) c^2_{13}
	\right]^2}
 \;,
\end{align}
\end{subequations}
where $s_{ij}\defeq\sin\theta_{ij}$, $c_{ij}\defeq \cos\theta_{ij}$ and
\begin{subequations}
\begin{align}
 a_0 & \defeq 
\left(\widetilde{m}_1 c ^2_{12}+\widetilde{m}_2 s ^2_{12}\right) c
 ^2_{13}+\ee^{2 \ii \deltaCP } m_3 s ^2_{13}\;,\\
 a_1 & \defeq 
 \left[\left(\ee^{2 \ii \deltaCP} s^2_{12}-c^2_{12} s^2_{13}\right) 
 \sin (2 \theta _{23})-\ee^{\ii \deltaCP } \cos (2 \theta _{23}) \sin (2 \theta_{12})
 s_{13}\right]
 \;,\\
 a_2 & \defeq 
 \left[\ee^{\ii \deltaCP} \cos(2 \theta_{23}) 
 \sin (2 \theta _{12}) s_{13}+\left(\ee^{2 \ii \deltaCP } c ^2_{12}-s^2_{12} 
 s^2_{13}\right) \sin (2 \theta _{23})\right]
 \;.
\end{align}
\end{subequations}
The invariants $I_{ij}$ depend on $m_1$, $m_2$, $\varphi_1$ and $\varphi_2$ only via the combinations $\widetilde{m}_1\defeq m_1\,\ee^{\ii\varphi_1}$ and $\widetilde{m}_2\defeq m_2\,\ee^{\ii\varphi_2}$.
As $I_{ij}$ are complex constants, they give rise to six relations. Based on these relations, one can infer, for instance, the scale dependence of all angles and phases from the running of the three mass eigenvalues.

While the expressions~\eqref{eq:RGinvariant} are lengthy, they have two important properties:
\begin{enumerate}
 \item they only depend on the physical parameters~\eqref{eq:LeptonMixingParameters};
 \item they are modular invariant.
\end{enumerate}
In models in which the Majorana neutrino masses are modular forms, the modular weights of the matrix elements of the light neutrino mass matrix are solely determined by the modular weights of the left-handed leptons $L_{i}$. 
Then the matrix element $M_{ij}(\tau)$ of the superpotential coupling matrix $M(\tau)$ has modular weight $k_{L_{i}} + k_{L_{j}}$.
The invariants~\eqref{eq:Definition_of_I_ij_in_MSSM}
must be modular functions of weight $2k_{L_i}+2k_{L_{j}}-2(k_{L_{i}} + k_{L_{j}})=0$ of the corresponding modular symmetry. 
This means that the $I_{ij}(\tau)$ can always be written as rational functions of corresponding so-called Hauptmodul of the corresponding modular symmetry~\cite{Sebbar:2002,Schultz:2015}.
The Hauptmodul for a given subgroup $G$ of $\text{SL}(2,\mathds{Z})$ is a modular function of weight 0 on $G$ which generates all the modular functions for this group $G$, and fulfills \Requirement{1} and \Requirement{2}.
The best-known example of this kind is the $j$-invariant $j(\tau)\defeq E_4^3(\tau)/\eta^{24}(\tau)$, which is the Hauptmodul for the full modular group $\mathrm{SL}(2,\mathds{Z})$. Here $E_{4}(\tau)$ is the Eisenstein series and $\eta(\tau)$ is the Dedekind eta function. 
Notice that these modular invariant functions have, as opposed to the modular forms, poles~\cite{ClassicalModularForms}, i.e.\ they fulfill \Requirement{1} \& \Requirement{2} but not \Requirement{3}.
Given these properties we have a significant amount of information on these physical observables directly from the theory of modular forms.
We will illustrate this crucial point in the following examples.

\subsection{Feruglio Model based on $\Gamma_3$}
\label{sec:Feruglio_Model_1}

Consider Model 1 from~\cite{Feruglio:2017spp}, which is based on finite modular group $\Gamma_3\cong A_4$.
The assignments of modular weights and representations for the matter fields are shown in \Cref{tab:Model1}.
\begin{table}[ht!]
\centering
\begin{tabular}{l*6{c}}
\toprule
field/coupling&$(E^c_1,E^c_2,E^c_3)$ &$L$ & $H_{u/d}$ & $\varphi_T$ & $Y^{(2)}_\rep{3}(\tau)$\\
\midrule 
$\mathrm{SU}(2)_{\text{L}}\times \mathrm{U}(1)_{\mathrm{Y}}$ &  $(\rep{1},1)$  & $(\rep{2},-1/2)$  &$(\rep{2},\pm 1/2)$ & $(\rep{1},0)$ & $(\rep{1},0)$ \\ 
$\Gamma_3\cong A_4$ &  $(\rep{1},~\rep{1''},~\rep{1'})$ & $\rep{3}$ & $\rep{1}$  &  $\rep{3}$ &  $\rep{3}$\\ 
$k_I$ &  $(2,2,2)$ & $1$  & $0$ &  $-3$ &  $-2$ \\ 
\bottomrule
\end{tabular}
\caption{Quantum numbers in Feruglio's Model 1.}
\label{tab:Model1}
\end{table}
The model contains a triplet flavon, $\varphi_T$, which only couples to the charged leptons. 
The effective neutrino masses depend only on the modular forms of weight $2$. 
In more detail, the relevant terms of the superpotential are given by
\begin{subequations}
\begin{align}
       \mathscr{W}_e&=\alpha E_1^c H_d\left(L \varphi_T\right)_\rep{1}+\beta E_2^c H_d\left(L \varphi_T\right)_{\rep{1}^{\prime}}+\gamma E_3^c H_d\left(L \varphi_T\right)_{\rep{1}^{\prime \prime}} \;,\\
       \mathscr{W}_\nu&=\frac{1}{\Lambda}\left(H_u\cdot L\, H_u\cdot L\,  Y^{(2)}_{\rep{3}}\right)_\rep{1}\;.
\end{align}
\end{subequations}
and there are no higher-order contributions to the effective neutrino mass operator in the superpotential.
The notation $(\dots)_{\rep{r}}$ indicates a contraction to the representation $\rep{r}$ of the finite group, i.e.\ $A_4$ in this case, and does not imply the $\text{SL}(2,\mathds{Z})$ representation explicitly.  
The flavon $\varphi_T$ is assumed to develop the \ac{VEV}
\begin{equation}
 \Braket{\varphi_T } = (u,0,0)\;.\label{eq:phiVEV}
\end{equation}
With the \ac{VEV} given in~\eqref{eq:phiVEV}, the charged lepton and neutrino mass matrices read
\begin{subequations}
\begin{align}
M_e&=u\, v_d\,\diag(\alpha,\beta,\gamma)
\;,\\
m_\nu(\tau,\bar{\tau})&=
\left(-\ii\tau+\ii\bar{\tau}\right)\frac{v_u^2}{\Lambda}\, 
\begin{pmatrix}
  2 Y_1(\tau) & -Y_2(\tau) & -Y_3(\tau) \\
  -Y_2(\tau) & 2 Y_3(\tau) & -Y_1(\tau) \\
  -Y_3(\tau) & -Y_1(\tau) & 2 Y_2(\tau)
  \end{pmatrix} \notag\\
  &\eqdef 
\left(-\ii\tau+\ii\bar{\tau}\right) v_u^2\,\begin{pmatrix}
  \kappa_{11} & \kappa_{12} & \kappa_{13} \\
  \kappa_{12} & \kappa_{22} & \kappa_{23} \\
  \kappa_{13} & \kappa_{23} & \kappa_{33}
  \end{pmatrix}\;.\label{eq:MnuFeruglioModels1}
\end{align}
\end{subequations}
Here, $Y_{1,2,3}$ are the components of modular forms triplet $Y^{(2)}_\rep{3}(\tau)$. 
Notice that $\alpha<\gamma<\beta$, and as a consequence $Y_2$ and $Y_3$ in~\eqref{eq:MnuFeruglioModels1} are swapped compared to~\cite[Equation~(38)]{Feruglio:2017spp}.

The invariants~\eqref{eq:Definition_of_I_ij_in_MSSM} are given by 
\begin{subequations}\label{eq:I_ij_Feruglio_Model_1}
\begin{align}
  I_{12}(\tau)&=4\frac{Y_1(\tau)\,Y_3(\tau)}{\bigl(Y_2(\tau)\bigr)^2}
  \;,\\
  I_{13}(\tau)&=4\frac{Y_1(\tau)\,Y_2(\tau)}{\bigl(Y_3(\tau)\bigr)^2}
  \;,\label{eq:I_ij_Feruglio_Model_1_I_13}\\
  I_{23}(\tau)&=4\frac{Y_2(\tau)\,Y_3(\tau)}{\bigl(Y_1(\tau)\bigr)^2}
  \;.\label{eq:I_ij_Feruglio_Model_1_I_23}
\end{align}
\end{subequations}
The invariants $I_{ij}(\tau)$ are products of ratios of two holomorphic modular forms, so they are meromorphic on the extended upper-half plane $\overline{\mathcal{H}}\defeq\mathcal{H}\cup \mathds{R}\cup\{\ii\infty\}$. 
They also transform as $A_4$ \rep{1}-plets.

Consider a modular invariant meromorphic function $\mathcal{I}(\tau)$. 
Modular invariance, as opposed to modular covariance, means that (cf.\ e.g.\ \cite{ClassicalModularForms})
\begin{enumerate}
  \item either $\mathcal{I}(\tau)$ is a $\tau$-independent constant, 
  \item or it has poles.
\end{enumerate}
Both cases are realized in the example at hand.
First of all, $I_{12}$ is a constant because the $Y_i$ satisfy the algebraic constraint\footnote{Two other interesting identities are $\left(Y_{\rep{3}}^{(2)}\,Y_{\rep{3}}^{(2)}\right)_{\rep{1}}=Y_1^2+2Y_2Y_3=E_4$ and $\left(Y_{\rep{3}}^{(2)}\,Y_{\rep{3}}^{(2)}\right)_{\rep{1'}}=Y_3^2+2Y_1Y_2=-12\eta^8$.}
\begin{equation}
\left(Y_{\rep{3}}^{(2)}\,Y_{\rep{3}}^{(2)}\right)_{\rep{1''}}=Y_2^2+2Y_1Y_3=0\;.\label{eq:Yconstraint}
\end{equation}
The latter follows from the fact that the modular form triplet $Y^{(2)}_\rep{3}(\tau)$ of weight 2 can be obtained from the tensor product of modular form doublet $Y^{(1)}_\rep{2}(\tau)\defeq \left(X_1(\tau),~X_2(\tau)\right)^{\transpose}$ of weight 1~\cite{Liu:2019khw},
\begin{equation}
Y^{(2)}_\rep{3}\defeq \begin{pmatrix}
Y_1 \\ Y_2 \\Y_3
\end{pmatrix}=\begin{pmatrix}
X_2^2 \\  \sqrt{2}X_1 X_2 \\ -X_1^2
\end{pmatrix}\;.
\end{equation}
Here, the modular forms $X_{1,2}(\tau)$ of weight 1 on $\Gamma(3)$ are given by
\begin{subequations}\label{eq:X1X2}
\begin{align}
X_1(\tau)&\defeq 3\sqrt{2}\frac{\eta^3(3\tau)}{\eta(\tau)}=3\sqrt{2}q^{1/3}(1+q+2q^2+2q^4+q^5+\dots)\;,\\
X_2(\tau)&\defeq -3\frac{\eta^3(3\tau)}{\eta(\tau)}-\frac{\eta^3(\tau/3)}{\eta(\tau)}=-1-6q-6q^3-6q^4-12q^7-\dots\; ,
\end{align}
\end{subequations}
where
\begin{equation}\label{eq:q_def}
  q=\ee^{2\pi\ii \tau}\;.
\end{equation}
Since the three $Y_i$ can be expressed in terms of the two $X_i$, the $Y_i$ are not algebraically independent, as manifest in the constraint in \Cref{eq:Yconstraint}.

Altogether we find, as expected, that the invariants can be expressed in terms of the Hauptmodul $j_3(\tau)\defeq\eta(\tau/3)^3/\eta(3\tau)^3$ of $\Gamma(3)$ as,
\begin{subequations}
\begin{align}
  I_{12}(\tau)&=-2\;,\label{eq:MIHO1}\\
  I_{13}(\tau)&=-2\left(1+\frac{1}{3}j_3(\tau)\right)^3\;,\label{eq:MIMO1}\\ 
  I_{23}(\tau)&=-\frac{32}{I_{13}}=\frac{16}{\left(1+\frac{1}{3}j_3(\tau)\right)^3}\;.\label{eq:MIMO2}
\end{align}  
\end{subequations}
\eqref{eq:MIMO1} and~\eqref{eq:MIMO2} are invariant under $\Gamma(3)$ and invariant under the full $\text{SL}(2,\mathds{Z})$ if one properly takes the transformation of the flavon $\varphi_T$ into account, cf.\ \Cref{sec:Feruglio_Model_1_modular_invariance_details}. 
Further, \Cref{eq:MIMO1,eq:MIMO2} imply that 
\begin{equation}\label{eq:Feruglio_Model_1_I_13_I_23}
  I_{13}\,I_{23}=-32\;.
\end{equation}
The $q$-expansions of $I_{13}$ and $I_{23}$ are given by
\begin{subequations}
\begin{align}
I_{13}&=-\frac{2}{27} q^{-1} -\frac{10}{9} - 4 q + \frac{152}{27}q^2 + 18 q^3 - 88 q^4 + \frac{2768}{27} q^5 + 216 q^6 +\dots \;,\\
I_{23}&= 432 q - 6480 q^2 + 73872 q^3 - 725328 q^4 + 6503328 q^5 - 54855792 q^6 +\dots\;.
\end{align}
\end{subequations}
It can be shown that $I_{13}$ has a singularity at  $\tau = \ii\infty$.  
Similarly, $I_{23}$ is singular at $\tau = \frac{-3+\ii\sqrt{3}}{6}$,  though it vanishes at  $\tau = \ii\infty$. 

The conditions in \Cref{eq:MIMO2,eq:MIMO1,eq:MIHO1} lead to robust phenomenological implications when relations in \Cref{eq:RGinvariant} are utilized where the invariants are expressed in terms of the physical mixing parameters~\eqref{eq:LeptonMixingParameters}. 
In particular, \Cref{eq:MIHO1} and \Cref{eq:Feruglio_Model_1_I_13_I_23} give rise to four constraints that are independent of the value of $\tau$.
Due to the form of the neutrino mass matrix in this model \Cref{eq:MnuFeruglioModels1}, there is a sum rule among the three physical neutrino masses~\cite{ChuliaCentelles:2022ogm,CentellesChulia:2023zhu},\footnote{Note that, unlike relations between the $I_{ij}$, this sum rule is not \ac{RG} invariant. 
Therefore, the numerical results presented in what follows are subject to corrections. These corrections can be readily computed in a given model~\cite{Antusch:2005gp}.} 
\begin{equation}
 m_3=\begin{dcases}
  m_2+m_1 & \text{for normal ordering (NO)}\;,\\
  m_2-m_1 & \text{for inverted ordering (IO)}\;.
 \end{dcases}  \label{eq:sumRule_A4}
\end{equation}
Given the sum rule, the three neutrino masses (and thus the absolute neutrino mass scale) are completely fixed by the two mass squared differences, $\Delta m^2_\mathrm{sol}$ and $\Delta m^2_\mathrm{atm}$, which have been determined from oscillation experiments. 
Furthermore, the mixing angles are also known from oscillation experiments~\cite{Esteban:2020cvm}.
We are thus left with three undetermined observables, namely the three \CP phases
\begin{equation}
\{ \deltaCP, \varphi_{1} , \varphi_{2}\}\;.
\label{eq:Phases}
\end{equation}
Hence we can use the invariants to predict the values of the \CP phases in this model.

\begin{figure}[h!]
  \centering
  \includegraphics[width=0.8\textwidth]{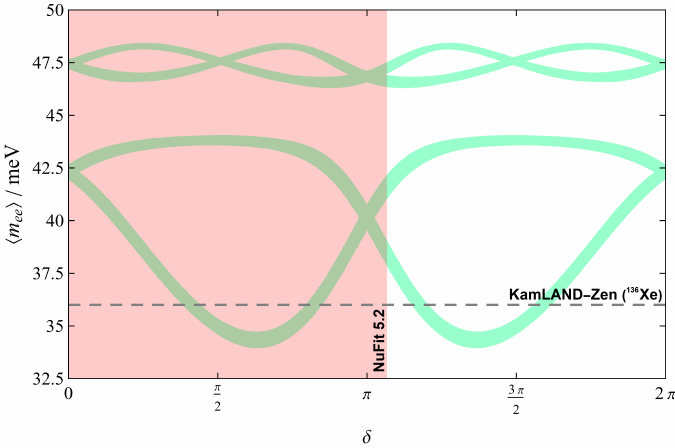} 
  \caption{This figure displays the correlation between the neutrinoless double $\beta$-decay matrix element $\Braket{m_{ee}}$ and the \CP phase $\deltaCP$ for \ac{IO} in the $\Gamma_3$ model,  considering only the invariant constraint $I_{12}=-2$. Note that this analysis is independent of $\tau$. The red-shaded region corresponds to the $3\sigma$ disfavored range of values for the Dirac phase $\deltaCP$ from the global fit~\cite{Esteban:2020cvm}, while the gray-dashed line represents the current experimental upper bound for $\langle m_{ee}\rangle$ from the KamLAND-Zen collaboration~\cite{KamLAND-Zen:2022tow}. The projected sensitivities of future experiments such as nEXO~\cite{nEXO:2017nam} can reach values for $\langle m_{ee}\rangle \sim 10\,\text{meV}$, and thus will probe the predictions of the invariant constraint $I_{12}=-2$ in this model.
  }
  \label{fig:DeltaVsMee}
\end{figure}

\Cref{eq:MIHO1} trivially fulfills requirements \Requirement{1}, \Requirement{2} \& \Requirement{3}.
It entails two constraints, $\re I_{12}=-2$ and $\im I_{12}=0$.
Therefore, for a given value of the Dirac \CP phase $\deltaCP$, we can predict the values of the Majorana phases $\varphi_1$ and $\varphi_2$. 
It is important to note that these predictions are independent of the value of $\tau$. With these predictions, one can then determine the neutrinoless double beta decay matrix element, $\Braket{m_{ee}}$, as shown in Figure \ref{fig:DeltaVsMee}. 
Given that only two out of the six conditions are utilized, the experimental best-fit values for the mixing parameters that have been used as our inputs in Figure \ref{fig:DeltaVsMee} may not be fully consistent with all constraints (in fact they are not, as we will discuss below). 
Nevertheless, it is interesting to see even with one invariant, it is already possible to significantly constrain the model.

After imposing \Cref{eq:MIMO2} \Cref{eq:MIHO1}, there is only one observable left undetermined.
If we impose \cref{eq:Feruglio_Model_1_I_13_I_23}, which entails two constraints, one for the real and one for the imaginary parts of $I_{13}\,I_{23}=-32$, the system is overconstrained. 
We have verified that we cannot impose the constraints~\eqref{eq:MIMO1} and~\eqref{eq:MIMO2} while still being consistent with data. 
These findings are consistent with the analyses in~\cite{Feruglio:2017spp,Criado:2018thu}, where it has been pointed out that one cannot accommodate all experimental data for the neutrino masses and mixings in this model. 
Note that we arrived at this conclusion without having to scan over $\tau$.
However, as discussed in~\cite{Criado:2018thu}, by adding one more parameter one can obtain a model which is remarkably consistent with the current experimental constraints.

\subsection{A model based on $\Gamma_5$}
\label{sec:A_5_Model}

We next consider a model based on $\Gamma_5\cong A_5$~\cite{Novichkov:2018nkm,Ding:2019xna,Criado:2018thu}.
The assignments of modular weights and representations for the matter fields are shown in \Cref{tab:A5Model}.
\begin{table}[ht!]
\centering
\begin{tabular}{l*7{c}}
\toprule
field/coupling&$E^c$ &$L$ & $H_{u/d}$ & $\chi$ & $\varphi$ & $Y^{(2)}_\rep{5}(\tau)$\\
\midrule 
$\text{SU}(2)_{\mathrm{L}}\times \mathrm{U}(1)_{\mathrm{Y}}$ &  $(\rep{1},1)$  & $(\rep{2},-1/2)$  &$(\rep{2},\pm 1/2)$ & $(\rep{1},0)$& $(\rep{1},0)$ & $(\rep{1},0)$ \\ 
$\Gamma_5\cong A_5$ &  $\rep{3}$ & $\rep{3}$ & $\rep{1}$  &  $\rep{1}$ &  $\rep{3}$&  $\rep{5}$\\ 
$k_I$ &  $2$ & $1$  & $0$ &  $-3/2$ &  $-3/2$&  $-2$ \\ 
\bottomrule
\end{tabular}
\caption{Quantum numbers in the $\Gamma_5$ model.}
\label{tab:A5Model}
\end{table}
This model introduces a singlet flavon, $\chi$, and an $A_5$ triplet flavon, $\varphi$, which only couple to the charged leptons. 
The effective neutrino masses depend only on the modular forms of weight $2$. 
More specifically, the relevant pieces of the superpotential are given by
\begin{subequations}
\begin{align}
\mathscr{W}_e&=\left[\alpha \left(E^c L \right)_\rep{1} \chi^2+\beta \left(E^c L \right)_\rep{1} (\varphi^2)_\rep{1}+\gamma \left(E^c L \right)_\rep{5}  (\varphi^2)_\rep{5} + \zeta \left(E^c L \right)_\rep{3}  (\chi\varphi)_\rep{3}\right]_{\rep{1}}\, H_d\;,\\
\mathscr{W}_\nu&=\frac{1}{\Lambda}\left(H_u\cdot L\, H_u\cdot L\,  Y^{(2)}_{\rep{5}}\right)_\rep{1}\;.
\end{align}
\end{subequations}
The symmetries of the model forbid higher-order contributions to the effective neutrino mass operator in the superpotential.
The flavons $\chi$ and $\varphi$ are assumed to attain the \acp{VEV}
\begin{subequations}\label{eq:A5phiVEV}
\begin{align}
  \Braket{\chi} &= v_\chi\;,\\  
  \Braket{\varphi} &= v_\varphi\,(1,0,0)\;.  
\end{align}  
\end{subequations}
With these \acp{VEV}, the charged lepton and neutrino mass matrices read
\begin{subequations}
\begin{align}
M_e&=v_d\,\begin{pmatrix}
\mu_e+4\gamma v_\varphi^2 & 0 & 0 \\ 0 & 0 & \mu_e-2\gamma v_\varphi^2+\zeta v_\chi v_\varphi \\ 0 & \mu_e-2\gamma v_\varphi^2-\zeta v_\chi v_\varphi & 0
\end{pmatrix} \;,\\
m_\nu(\tau,\bar{\tau})&=
\left(-\ii\tau+\ii\bar{\tau}\right)\frac{v_u^2}{\Lambda}\, 
\begin{pmatrix}
  2 Y_1(\tau) & -\sqrt{3}Y_5(\tau) & -\sqrt{3}Y_2(\tau) \\
  -\sqrt{3}Y_5(\tau) & \sqrt{6} Y_4(\tau) & -Y_1(\tau) \\
  -\sqrt{3}Y_2(\tau) & -Y_1(\tau) & \sqrt{6} Y_3(\tau)
  \end{pmatrix} 
  \notag\\
 &\eqdef 
\left(-\ii\tau+\ii\bar{\tau}\right)
 v_u^2\,\begin{pmatrix}
  \kappa_{11} & \kappa_{12} & \kappa_{13} \\
  \kappa_{12} & \kappa_{22} & \kappa_{23} \\
  \kappa_{13} & \kappa_{23} & \kappa_{33}
  \end{pmatrix}\;,\label{eq:MnuA5Model}
\end{align}
\end{subequations}
where $\mu_e\defeq\alpha\, v_\chi^2+\beta\, v_\varphi^2$.
Here, $Y_{1,2,3,4,5}$ are the components of modular forms quintuplet $Y^{(2)}_\rep{5}(\tau)$. 
They are not algebraically independent. 
In fact, each of them can be written as a homogeneous polynomial of 10 degrees in two basic modular forms of weight $\nicefrac{1}{5}$, $F_{1}(\tau)$ and $F_{2}(\tau)$,
\begin{equation}\newcommand*{\myVP}{\vphantom{\sqrt{6} F_1^4(\tau)F_2(\tau)\left(F^5_1(\tau)+7F_2^5(\tau)\right)}}
\label{eq:MFquiintuplet}
Y^{(2)}_\rep{5}(\tau)\defeq\begin{pmatrix}
  \myVP Y_1(\tau) \\ \myVP Y_2(\tau) \\ \myVP Y_3(\tau) \\ \myVP Y_4(\tau) \\ \myVP Y_5(\tau)
\end{pmatrix} = \begin{pmatrix}
-F_1^{10}(\tau)-F_2^{10}(\tau) \\ \sqrt{6} F_1^4(\tau)F_2(\tau)\left(F^5_1(\tau)+7F_2^5(\tau)\right) \\ 
\sqrt{6} F_1^3(\tau)F^2_2(\tau)\left(3F^5_1(\tau)-4F_2^5(\tau)\right) \\
\sqrt{6} F_1^2(\tau)F^3_2(\tau)\left(4F^5_1(\tau)+3F_2^5(\tau)\right) \\ 
-\sqrt{6} F_1(\tau)F^4_2(\tau)\left(-7F^5_1(\tau)+F_2^5(\tau)\right)
\end{pmatrix} \;,
\end{equation}
where~\cite{Ibukiyama:2000mf,Yao:2020zml}
\begin{subequations}\label{eq:A5_F_i_def}
\begin{align}
F_1(\tau)&\defeq
\ee^{-\pi\ii/10}\dfrac{\vartheta_{(\nicefrac{1}{10},\nicefrac{1}{2})}(5\tau)}{\bigl(\eta(\tau)\bigr)^{3/5}}
=\dfrac{\sum\limits_{p\in\Z{}}(-1)^{p}q^{(5p^2+p)/2}}{\prod\limits_{n=1}^{\infty}(1-q^n)^{3/5}}
\;,\\
F_2(\tau)&\defeq
\ee^{-3\pi\ii/10}\dfrac{\vartheta_{(\nicefrac{3}{10},\nicefrac{1}{2})}(5\tau)}{\bigl(\eta(\tau)\bigr)^{3/5}}
= \dfrac{q^{1/5}\sum\limits_{p\in\Z{}}(-1)^{p}q^{(5p^2+3p)/2}}{\prod\limits_{n=1}^{\infty}(1-q^n)^{3/5}} \; ,
\end{align}
\end{subequations}
with $q$ from~\eqref{eq:q_def}, $\eta(\tau)$ being the Dedekind $\eta$-function defined before, and the $\vartheta$-constants given by
\begin{equation}
  \vartheta_{(\mu,\nu)}=\sum\limits_{m\in\mathds{Z}}
  \exp\left\{2\pi\ii\left[\frac{1}{2}(m+\mu)^2\tau+(m+\mu)\,\nu\right]\right\}\;.
\end{equation}

The Hermitean combination $M_e^\dagger M_e$ is diagonal, and the three charged lepton masses can be obtained by adjusting the free parameters $\alpha$, $\beta$, $\gamma$, and $\zeta$. 
As before, we work in the basis in which the charged lepton Yukawa coupling is diagonal and the diagonal entries fulfill $(M_e^\dagger M_e)_{11}<(M_e^\dagger M_e)_{22}<(M_e^\dagger M_e)_{33}$. 
The best-fit value of modulus $\tau$ is also close to the critical point $\ii$,
\begin{equation}
\langle\tau\rangle=-0.0219308+0.994295\ii \;.
\end{equation}
These six real input parameters lead to the following neutrino mass and mixing parameters, as shown in \Cref{tab:A5_predictions}.\footnote{Note that the precision with which we present predictions of the model is misleading in that we do not have sufficient theoretical control over the model. These are ``mathematical predictions'' which allow other research groups to cross-check our results. As discussed around \Cref{eq:MinimalK}, there are limitations, and generally it is nontrivial to make the theoretical error bars smaller than the experimental ones, see~\cite{Almumin:2022rml} for a more detailed discussion.\label{ftn:Precision}}
\begin{table}[t]
  \centering 
  \begin{tabular}{l}
    \toprule 
    $\begin{aligned}
    \text{masses} && m_1&=49.2354\,\text{meV} & m_2&=49.9912\,\text{meV} & m_3&=1.05795\,\text{meV}\\
    \text{angles} && \sin^2\theta_{12}&=0.283166 & \sin^2\theta_{13}&=0.0345963 & \sin^2 \theta_{23}&=0.795443 \\
    \text{phases} && \deltaCP/\pi&=1.39102 & \varphi_1/\pi&=0.141799 & \varphi_2/\pi&=0.517899\\
    \end{aligned}$\\
    \bottomrule   
  \end{tabular}
  \caption{Predictions from the best-fit point of the $\Gamma_5$ model. 
  The neutrino masses are predicted to be of inverted ordering. 
  $\sin^2\theta_{13}$ deviates from the central value by about $20\sigma$  and $\sin^2\theta_{23}$ deviates from the central value by about $10\sigma$.
  The remaining observables fall within the $3\sigma$ ranges of the experimental data (cf.\ NuFIT 5.2 without SK~\cite{Esteban:2020cvm}).}
  \label{tab:A5_predictions}
\end{table}
The ratio of mass squared differences is given by
\begin{equation}
  \frac{\Delta m_{\text{sol}}^2}{\left|\Delta m_{\text{atm}}^2\right|}=0.03\;.
\end{equation}

While we have not found a set of input values that give rise to predictions that are consistent with all experimental data, it would still be interesting, as in the case of the $\Gamma_{3}$ Model, to see how robust relations could arise by considering the invariants. 
The \ac{RG} invariants emerging from the neutrino mass matrix~\eqref{eq:MnuA5Model} are given by
\begin{subequations}
\label{eq:I_ij_A5Model}
\begin{align}
I_{12}&=\frac{2\sqrt{6}}{3}\frac{Y_1(\tau)Y_4(\tau)}{Y^2_5(\tau)}\;,\\
I_{13}&=\frac{2\sqrt{6}}{3}\frac{Y_1(\tau)Y_3(\tau)}{Y^2_2(\tau)}\;, \\
I_{23}&=6\frac{Y_3(\tau)Y_4(\tau)}{Y^2_1(\tau)}\;.
\end{align}
\end{subequations}
These \ac{RG} invariants are meromorphic modular functions on $\Gamma(5)$ rather than $\text{SL}(2,\mathds{Z})$. 
As a consequence, they can be written as rational polynomials of the Hauptmodul $j_5(\tau)$ of $\Gamma(5)$.
Further, the $q$-expansions of $I_{12}$, $I_{13}$ and $I_{23}$ are given by
\begin{subequations}
\label{eq:qexpansion_I_ij_A5Model}
\begin{align}
I_{12}&=-\frac{8}{147} q^{-1} -\frac{338}{1029} - \frac{4420}{7203} q + \frac{260}{16807}q^2 + \frac{125120}{352947} q^3 - \frac{856444}{2470629} q^4 +\dots \;,\\
I_{13}&=-2 +\frac{92}{3}q - \frac{1460}{3} q^2 + 6960 q^3 - \frac{284260}{3} q^4 +1248060 q^5 +\dots \;,\\
I_{23}&= 432 q - 2412 q^2 + 7704 q^3 - 6876 q^4 - 93240 q^5+\dots\;.
\end{align}
\end{subequations}
Unlike in the $\Gamma_{3}$ model discussed in \Cref{sec:Feruglio_Model_1}, none of the $I_{ij}$ is a constant. 
In particular, $I_{12}$ has a singularity at $\tau = \ii\infty$, $I_{13}$ is singular at $\tau = \frac{1+0.767664\ii}{2}$, and $I_{23}$ is singular at $\tau = \frac{2+\ii}{5}$.

Since $Y_{1,2,3,4,5}$ can be expressed in terms of two building blocks, cf.\ \Cref{eq:MFquiintuplet}, these three invariants $I_{ij}$ are also the rational polynomials of the same building blocks. 

The way the algebraic relations between the invariants get specified is not unique. 
In what follows, we show one possible way,
\begin{subequations}
\label{eq:relations_A5Model}
\begin{align}
0&= 4+18 I_{12}+18 I_{13}+9 I_{12} I_{13}+I_{12} I_{13} I_{23}\;,\label{eq:independentRelations_A5Model_a}  \\
0&= 8+12 I_{12}-108 I_{12}^2+12 I_{13}+414 I_{12} I_{13}+108 I_{12}^2 I_{13}-108 I_{13}^2+108 I_{12} I_{13}^2+81 I_{12}^2 I_{13}^2 \nonumber\\
 &\quad{}-I_{12}^2 I_{23}-I_{13}^2 I_{23} 
 \;.\label{eq:independentRelations_A5Model_c}
\end{align}
\end{subequations}
These relations are richer than the corresponding constraint~\eqref{eq:Yconstraint} in the $\Gamma_{3}$ model of \Cref{sec:Feruglio_Model_1}.
However, they have the same qualitative virtue as their pendants of \Cref{sec:Feruglio_Model_1}: they allow us to derive constraints on the observables of the model.
Interestingly, \Cref{eq:relations_A5Model} are invariant under the exchange $I_{12}\leftrightarrow I_{13}$.
At the level of the functions of observables~\eqref{eq:RGinvariant}, this transformation is equivalent to $\theta_{23}\mapsto \theta_{23}+\pi/2$. 
This transformation is also known as $\mu\leftrightarrow\tau$ symmetry or $2-3$ symmetry~\cite{Lam:2001fb} (see e.g.~\cite{Xing:2015fdg} for a review), and has been considered in the context of modular flavor symmetries in~\cite{Ding:2023ynd}. 
It is, therefore, worthwhile to explore the ``fixed point'' of this exchange symmetry, i.e.\ make the ansatz that
\begin{equation}
  I_{12}=I_{13}\;.\label{eq:Gamma5_fixed_point} 
\end{equation}
$I_{12}$ and $I_{13}$ depend on $F_1$ and $F_2$, and in the limit in which either of the $F_i$ becomes zero at least one of the invariants becomes undefined. 
Therefore, we can assume $F_1\ne0$, and define $z\defeq F_2/F_1$.

\paragraph{(i) $F_1^{10}(\tau)= - F_2^{10}(\tau)$.}
Let us first the special case in which $F_1^{10}(\tau)= - F_2^{10}(\tau)$. In this case, $Y_1(\tau)=0$ and consequently 
\begin{equation}\label{eq:A5_invariant_fixed_point_relations2}
  I_{12}=I_{13}=0\quad\text{and}\quad I_{23}=\infty\;.
\end{equation}
In this case, the ratio $z$ can take any of the 10 values
\begin{equation}
  z_n^{(\mathrm{I})}=\ee^{\pi\ii (2n+1)/10}  \; , \quad n = 0, \dots, 9 \;.
\end{equation}
These solutions predict $\sin^2\theta_{23}=\nicefrac{1}{2}$, and $\sin^2\theta_{13} = \nicefrac{1}{3}$. Furthermore, the two larger mass eigenvalues are predicted to be degenerate with the sum rule $m_1-m_2=m_3=0$ (and thus \ac{IO}).

For instance, $z_9^{(\mathrm{I})}=\ee^{19\pi\ii/10}$ corresponds to $\tau=-2/5+\ii/5$. Note that $\tau=-2/5+\ii/5$ is a fixed point under the stabilizer $\mathds{Z}_2=\{1, ST^2S(ST^2)^{-1}\}$. 
At this fixed point, the neutrino mass matrix also has a generalized $\mu\leftrightarrow\tau$ symmetry,
\begin{equation}
\label{eq:mu-tau@z9}
 m_\nu= -\bigl(\rho_{\rep{3}}(ST^2S(ST^2)^{-1})\bigr)^{\transpose}\, m_\nu\, \rho_{\rep{3}}(ST^2S(ST^2)^{-1})\;, 
\end{equation}
where 
\begin{equation}
\rho_{\rep{3}}\bigl(ST^2S(ST^2)^{-1}\bigr)=-\begin{pmatrix}
  1 & 0 & 0 \\
  0 & 0 & \ee^{4\pi\ii /5} \\
 0 &  \ee^{-4\pi\ii n/5} & 0 
 \end{pmatrix}\;.
\end{equation}
and the $-1$ in \Cref{eq:mu-tau@z9} comes from the automorphy factor $(c\tau+d)^2=-1$. 

\paragraph{(ii) $F_1^{10}(\tau)\ne- F_2^{10}(\tau)$.}
Now  consider the more general case, $F_1^{10}\ne F_2^{10}$, i.e.\ $I_{12},I_{13}\ne0$.
The fixed point relation~\eqref{eq:Gamma5_fixed_point} can be traded for a constraint on $z$, which turns out to be a polynomial of degree 10. 
The 10 solutions are given by 
\begin{equation}
\label{eq:Gamma5_fixed_point_z_n}
 z_n^{(\mathrm{II})}=\frac{\sqrt{5}-(-1)^n}{2}\,\ee^{2\pi\ii n/10} \; , \quad n = 0, \dots, 9  \;. 
\end{equation} 
We can express the \ac{RG} invariants in terms of $z$,
\begin{subequations}
\label{eq:Gamma5_fixed_point_I_ij}
\begin{align}
  I_{12}&=-\frac{2 \left(3 z^5+4\right) \left(z^{10}+1\right)}{3 z^5 \left(z^5-7\right)^2}\;,\\
  I_{13}&=\frac{2 \left(4 z^5-3\right)
   \left(z^{10}+1\right)}{3 \left(7 z^5+1\right)^2}\;,\\
  I_{23}&=-\frac{36 z^5 \left(3 z^5+4\right) \left(4
   z^5-3\right)}{\left(z^{10}+1\right)^2}\;.
\end{align}  
\end{subequations}
Clearly, the $I_{ij}$ are rational functions of $z^5$. Since $z^5$ is real, $I_{ij}$ are real.
In fact, for all 10 solutions in~\eqref{eq:Gamma5_fixed_point_z_n},
\begin{equation}\label{eq:A5_invariant_fixed_point_relations}
 I_{12}=I_{13}=-\frac{2}{3}\quad\text{and}\quad I_{23}=36\;.
\end{equation} 
As one would expect, $\theta_{23}$ is maximal, i.e.\ $\sin^2\theta_{23}=\nicefrac{1}{2}$, and $\sin^2\theta_{13} = \nicefrac{1}{5}$. 
The size of the mass eigenvalues depends only on whether $n$ is even or odd, i.e.\ $|z_n^{(\mathrm{II})}|$.
For each $z_n^{(\mathrm{II})}$, there exist $\tau_n^{(\mathrm{II})}$ such that $r_{21}\bigl(\tau_n^{(\mathrm{II})}\bigr)\defeq F_2\bigl(\tau_n^{(\mathrm{II})}\bigr)/F_1\bigl(\tau_n^{(\mathrm{II})}\bigr)\simeq z_n^{(\mathrm{II})}$. 
All even (odd) $n$, $\tau_n^{(\mathrm{II})}$ can be obtained from $\tau_0^{(\mathrm{II})}$ ($\tau_1^{(\mathrm{II})}$) via $\text{SL}(2,\mathds{Z})$ transformations, i.e.\ for $m\in\mathds{Z}_5$
\begin{subequations}\label{eq:A5_case_II_z_n_equations}
\begin{align}
 r_{21}\bigl(\tau_{2m}^{(\mathrm{II})}\bigr)&\simeq z_{2m}^{(\mathrm{II})}\;,\\
 r_{21}\bigl(\tau_{5+2m}^{(\mathrm{II})}\bigr)&\simeq z_{5+2m}^{(\mathrm{II})}\;.  
\end{align}  
\end{subequations}
While there are no exact analytic solutions, the following $\tau$ values
\begin{subequations}
\begin{alignat}{2}
 \tau_{2m}^{(\mathrm{II})}&=\tau_0^{(\mathrm{II})}+m\;,\quad&\text{where } \tau_0^{(\mathrm{II})}&\simeq \varepsilon\ii\;,\\
 \tau_{5+2m}^{(\mathrm{II})}&= \tau_5^{(\mathrm{II})}+m\;,\quad&\text{where } \tau_5^{(\mathrm{II})}&\simeq
 2.5+\varepsilon'\ii\;,
\end{alignat}  
\end{subequations}
 with $0<\varepsilon,\varepsilon'\ll1$, solve~\eqref{eq:A5_case_II_z_n_equations} almost perfectly.
The appearance of the relative phases between $z_n^{(\mathrm{II})}$ can be seen easily from the definition of the $F_i$ in~\eqref{eq:A5_F_i_def}.
Note also that all $\tau_{2m}^{(\mathrm{II})}$ are related via $\text{SL}(2,\mathds{Z})$ but not $\Gamma_{5}$ transformations, and likewise for $\tau_{2m+1}^{(\mathrm{II})}$.

However, the solutions in~\eqref{eq:Gamma5_fixed_point_z_n} also predict the unrealistic relation $m_1=m_2$ along with the sum rule $m_3=m_1+m_2$ (and thus NO) and, as a consequence of~\eqref{eq:Gamma5_fixed_point_I_ij}, vanishing phases. 
The sum rule implies a continuous symmetry of the neutrino mass matrix,
\begin{equation}
  R(\theta)\cdot m_\nu\cdot \bigl(R(\theta)\bigr)^T=m_\nu\;,  
\end{equation}
where $R(\theta)\defeq U_\mathrm{PMNS}^\dagger\cdot R_{3}(\theta)\cdot U_\mathrm{PMNS}^{\transpose}$ with $R_{3}(\theta)$ being a rotation in the $1-2$ plane.

Furthermore, the predicted relations for the masses and mixing angles are a consequence of an approximate discrete symmetry of the neutrino mass matrix
\begin{equation}\label{eq:A5_case_II_m_nu_symmetry}
  m_\nu= \bigl(U_{\rep{3}}(n)\bigr)^{\transpose}\, m_\nu\, U_{\rep{3}}(n)\;, 
\end{equation}
where the Hermitean unitary matrix $U_{\rep{3}}(n)$ squares to unity and is given by
\begin{equation}\label{eq:A5_case_II_U_3_n}
  U_{\rep{3}}(n)=
 -{}\begin{pmatrix}
  1 & 0 & 0 \\
  0 & 0 & \ee^{2\pi\ii n/5} \\
 0 &  \ee^{-2\pi\ii n/5} & 0 
 \end{pmatrix} \;.
\end{equation}
This transformation can be regarded as a $\mathds{Z}_2$ transformation of the \rep{5}-plet,
\begin{equation}
  \begin{pmatrix}
    \vphantom{\ee^{2\pi\ii n/5}}Y_1\\ \vphantom{\ee^{2\pi\ii n/5}}Y_2\\ \vphantom{\ee^{2\pi\ii n/5}}Y_3\\ \vphantom{\ee^{2\pi\ii n/5}}Y_4\\ \vphantom{\ee^{2\pi\ii n/5}}Y_5
   \end{pmatrix} 
 \mapsto 
 \begin{pmatrix}
  \vphantom{\ee^{2\pi\ii n/5}}1 & 0 & 0 & 0 & 0 \\
 0 & 0 & 0 & 0 & \ee^{2\pi\ii n/5} \\
 0 & 0 & 0 & \ee^{4\pi\ii n/5}  & 0 \\
 0 & 0 & \ee^{-4\pi\ii n/5}  & 0 & 0 \\
 0 & \ee^{-2\pi\ii n/5}  & 0 & 0 & 0 
 \end{pmatrix} 
 \cdot 
 \begin{pmatrix}
  \vphantom{\ee^{2\pi\ii n/5}}Y_1\\ \vphantom{\ee^{2\pi\ii n/5}}Y_2\\ \vphantom{\ee^{2\pi\ii n/5}}Y_3\\ \vphantom{\ee^{2\pi\ii n/5}}Y_4\\ \vphantom{\ee^{2\pi\ii n/5}}Y_5
 \end{pmatrix} \eqdef U_{\rep{5}}(n)\cdot 
 \begin{pmatrix}
  \vphantom{\ee^{2\pi\ii n/5}}Y_1\\ \vphantom{\ee^{2\pi\ii n/5}}Y_2\\ \vphantom{\ee^{2\pi\ii n/5}}Y_3\\ \vphantom{\ee^{2\pi\ii n/5}}Y_4\\ \vphantom{\ee^{2\pi\ii n/5}}Y_5
 \end{pmatrix} \;.
\end{equation}
We emphasize that none of the symmetries~\eqref{eq:A5_case_II_m_nu_symmetry} are exact symmetries of the action, they are symmetries of neutrino mass matrix at the fixed point of the syzygies~\eqref{eq:relations_A5Model}.
However, in this setup having symmetries of the neutrino mass matrix, as opposed to symmetries of the action, comes at a price: the modular forms become very large, their absolute values can exceed 100. In the context of bottom-up model building this can be acceptable because there is no a priori normalization of the modular forms, i.e.\ we can always multiply them with a small constant. 
It is to be noted that symmetries of mass matrices have been discussed in the literature. 
However, to the best of our knowledge, these symmetries have been imposed in a rather ad hoc fashion in the sense that there is no model realization for these previous examples. 
We speculate that the type of model construction considered in this Letter may provide a consistent framework from which symmetries of mass matrices can arise dynamically. 
We will investigate this aspect further in a subsequent work.

\section{Discussion}
\label{sec:Discussion}

As we have seen, in modular invariant models of flavor, it is possible to relate certain meromorphic modular invariant functions to physical observables. 
In some cases, such as~\eqref{eq:MIHO1} and~\eqref{eq:Feruglio_Model_1_I_13_I_23}, one even obtains modular invariant holomorphic observables, where a combination of observables conspires to become an integer, independent of the renormalization scale.
We have shown that useful information and phenomenological constraints can be extracted from these relations, which, due to \ac{RG} invariance, can be directly, modulo the limitations discussed around \Cref{eq:MinimalK}, applied to observables measured in experiments.
It will also be interesting to apply our discussion to the quark sector, where invariants were obtained from different considerations~\cite{Wang:2021wdq,Bento:2023owf}. 

The fact that these observables conspire to be integers may be regarded as a hint towards a topological origin of these relations. 
In the effective theory approach, it is not obvious how to substantiate such speculations. 
However, it has been known long before modular invariance was used in bottom-up model building that the couplings in string compactifications are modular forms~\cite[cf.\ the discussion around Equation (19)]{Quevedo:1996sv}.
Earlier work~\cite{Lauer:1989ax,Chun:1989se,Lauer:1990tm} and more recent analyses 
\cite{Baur:2019kwi,Baur:2019iai,Nilles:2020nnc,Nilles:2020kgo,Nilles:2020tdp,Baur:2020jwc,Nilles:2020gvu,Baur:2020yjl,Baur:2021mtl,Nilles:2021ouu,Nilles:2021glx,Baur:2021bly,Baur:2022hma,Knapp-Perez:2023nty} explore the stringy origin of these couplings. 
It will be interesting to see whether the above-mentioned integers, which can be directly related to experimental observation as we have shown, play a special role in stringy completions of the \ac{SM}.
It is tempting to speculate that this may provide us with a direct relation between experimental measurements and properties of the compact dimensions. 

Obviously, this is not the first time in which holomorphy (\Requirement{2}) and modular invariance (\Requirement{1}) is used to make firm physical predictions. 
In particular, the celebrated Seiberg--Witten theory~\cite{Seiberg:1994rs,Seiberg:1994aj} makes use of these  concepts to solve gauge theories with $\mathcal{N}=2$ \ac{SUSY}. 
However, our discussion shows, in the framework of modular flavor symmetries, \Requirement{1}, \Requirement{2}, and in some instances \Requirement{3} govern certain combinations of real-world observables. 
As we discussed, these combinations are \ac{RG} invariant to all orders within $\mathcal{N}=1$ \ac{SUSY}, and their poles and zeros are \ac{RG} invariant even without \ac{SUSY}. 
Let us reiterate that these conclusions are generally valid only under the assumption that the K\"ahler potential attains its minimal form~\eqref{eq:MinimalK} at some scale. 
It will, therefore, be interesting to find alternatives to \cite{Chen:2021prl} allowing us to control the K\"ahler potential.
Likewise, the discussion of modular invariant holomorphic observables for non-minimal K\"ahler potentials are left to future work.

\clearpage
\section{Summary}
\label{sec:Summary}

We have pointed out that in modular invariant models of flavor, certain combinations of couplings give rise to modular invariant meromorphic and even holomorphic physical observables. These objects are highly constrained by their symmetries and properties, \ac{RG} invariant, and, at the same time, composed solely of quantities that can be measured experimentally. They carry a lot of information, and allow us to draw immediate, important, and robust conclusions on the model without the need  to perform scans of the parameter space. 
In addition, symmetry relations among the invariants exist for certain modular symmetries, as illustrated in the $\Gamma_{5}$ model studies in this Letter. Fundamentally they are symmetries of the fixed points and can correspond to phenomenologically relevant ones, such as the $\mu-\tau$ symmetry.

More importantly, to the best of our knowledge, these are the first examples in which physical observables are given by modular invariant functions. 
This Letter is only the start of exploiting their properties to obtain better theoretical control of model predictions.

\section*{Acknowledgments}

We would like to thank Yuri Shirman for useful discussions.
One of us (M.R.) would like to thank Manfred Lindner and Michael A.\ Schmidt for useful discussions and collaboration on the \ac{RG} invariants, which we recap in \Cref{sec:RGinvariants}. O.M.\ thanks the Department of Physics and Astronomy at UC, Irvine, for their hospitality during his visit.
The work of M.-C.C., X.-G.L.\ and M.R.\ is supported by the National Science Foundation, under Grant No.\ PHY-1915005. 
The work of M.-C.C.\ and M.R.\ is also supported by UC-MEXUS-CONACyT grant No.\ CN-20-38.
O.M.\ is supported by the Spanish grants PID2020-113775GB-I00 (AEI/10.13039/501100011033), Prometeo CIPROM/2021/054 (Generalitat Valenciana), Programa Santiago Grisolía (No.\ GRISOLIA/2020/025) and the grant for research visits abroad CIBEFP/2022/63.

\appendix

\section{Renormalization-group invariant expressions}
\label{sec:RGinvariants}

Let us now study the \ac{RG} evolution for the effective neutrino mass operator. 
The structure of the \ac{RGE} in the \ac{SM}, \acp{2HDM} and the \ac{MSSM} is
\begin{equation}\label{eq:RGEkappa}
 16\pi^2\,\frac{\dd}{\dd t}\kappa=P^{\transpose}\,\kappa+\kappa\,P+\alpha\,\kappa\;,
\end{equation}
where at one-loop $P=C_e\,Y_e^\dagger Y_e$ with $Y_e$ being the charged lepton Yukawa matrix, and $t=\ln(\mu/\mu_0)$. 
As usual, $\mu$ denotes the renormalization scale and $\mu_0$ a reference scale. 
The coefficients $C_e$ are $C_e=-3/2$ in the \ac{SM}~\cite{Antusch:2001ck} and two-Higgs models~\cite{Antusch:2001vn}, and $C_e=1$
in the \ac{MSSM}~\cite{Chankowski:1993tx,Babu:1993qv}. 
In the basis where $P$ is diagonal it is easy to see that 
\begin{equation}\label{eq:RGE_kappa}
 \Delta \kappa_{ij}
 =
 \frac{\Delta t}{16\pi^2}\,\kappa_{ij}\,\left(P_{ii}+P_{jj}+\alpha\right)
 \;,
\end{equation}
where no summation over $i,j$ is implied. 
It has been pointed out in~\cite{Chang:2002yr} that certain ratios of entries of $\kappa$ do not depend on the renormalization scale,
\begin{align}
 I_{ij} & = \frac{\kappa_{ii}\,\kappa_{jj}}{\kappa_{ij}^2}
 \quad(i\ne j)\;. 
\end{align}
In the \ac{MSSM}, this can be understood from the non-renormalization theorem. 
Here, only the wave-function renormalization constants are scale-dependent, and this dependence precisely cancels in the above expressions~\cite{Haba:1999ca}. 
It has been noted in~\cite{Chang:2002yr} that this statement also applies to the  non-supersymmetric \ac{SM} at the one-loop level.

The scale invariance of $I_{ij}$ is due to the fact that the renormalizable couplings in the \ac{SM} have a larger global symmetry.
Specifically, the lepton sector has global lepton family number symmetries.
Therefore, in the basis in which the charged lepton mass matrix is diagonal, corrections that multiply the effective neutrino mass operator will be diagonal as well.
As a result, 
\begin{equation}\label{eq:full_RGE}
  \frac{\dd}{\dd t}\kappa=\widetilde{P}\,\kappa\,\widetilde{Q}^{\transpose}+\widetilde{Q}\,\kappa\,\widetilde{P}^{\transpose}+\widetilde{\alpha}\,\kappa\;, 
\end{equation}
where $\widetilde{P}$, $\widetilde{Q}$ and $\widetilde{\alpha}$ are composed of the renormalizable couplings of the theory and diagonal,
\begin{subequations}
\begin{align}
  \widetilde{P}&=\diag(\widetilde{P}_1,\widetilde{P}_2,\widetilde{P}_3)\;,\\
  \widetilde{Q}&=\diag(\widetilde{Q}_1,\widetilde{Q}_2,\widetilde{Q}_3)\;.  
\end{align}  
\end{subequations} 
At 1-loop, $\widetilde{P}=\frac{1}{16\pi^2}P$, $\widetilde{Q}=\mathds{1}$, and $\widetilde{\alpha}=\frac{1}{16\pi^2}\alpha$. 
\Cref{eq:full_RGE} implies that
\begin{equation}
  \dot\kappa_{ij}
  =\kappa_{ij}\bigl(\widetilde{P}_{i}\,\widetilde{Q}_{j}+\widetilde{P}_{j}\,\widetilde{Q}_{i}+\widetilde{\alpha}\bigr)\;,
\end{equation}
where no summation over $i$ or $j$ is implied.
This means that 
\begin{align}
 \frac{\dd}{\dd t}I_{ij} 
 &=\frac{\dot\kappa_{ii}\,\kappa_{jj}}{\kappa_{ij}^2}
 +\frac{\kappa_{ii}\,\dot\kappa_{jj}}{\kappa_{ij}^2}
 -2\frac{\kappa_{ii}\,\kappa_{jj}}{\kappa_{ij}^3}\,\dot\kappa_{ij}
 \notag\\
 &=2\bigl(\widetilde{P}_{i}-\widetilde{P}_{j}\bigr)\,\bigl(\widetilde{Q}_{i}-\widetilde{Q}_{j}\bigr)\,I_{ij}\;.
\end{align}
This has two immediate consequences:
\begin{enumerate}
 \item At 1-loop, where $\widetilde{Q}_{i}=1$ for all $i$, $I_{ij}$ are \ac{RG} invariant.
 \item Zeros and poles of $I_{ij}$ remain zeros and poles at \emph{all orders}.   
\end{enumerate}
In particular, in the basis in which $P$ is diagonal one can write the scale-dependent neutrino mass operator as
\begin{equation}
 \kappa(\mu) =
 \begin{pmatrix}
  z_1(\mu)\,z_1(\mu) & z_1(\mu)\,z_2(\mu)\,I_{12}^{-\nicefrac{1}{2}} &
  z_1(\mu)\,z_3(\mu)\,I_{13}^{-\nicefrac{1}{2}}\\
  z_2(\mu)\,z_1(\mu)\,I_{12}^{-\nicefrac{1}{2}} & z_2(\mu)\,z_2(\mu) &
  z_2(\mu)\,z_3(\mu)\,I_{23}^{-\nicefrac{1}{2}}\\
  z_3(\mu)\,z_1(\mu)\,I_{13}^{-\nicefrac{1}{2}} & z_3(\mu)\,z_2(\mu)\,I_{23}^{-\nicefrac{1}{2}} & z_3(\mu)\,z_3(\mu)
 \end{pmatrix}
\end{equation}
as long as $\kappa$ does not have zeros. 
As indicated, only the $z_i=\sqrt{\kappa_{ii}}$ are subject to \ac{RG} evolution. 
In slightly more detail, only the absolute values of the $z_i$ depend on the scale while their phases remain invariant. 
If one or more entries of $\kappa$ are zeros, then our discussion shows that these entries remain zero at all scales in the perturbative \ac{EFT} description. 
Zeros of the diagonal (off-diagonal) entries of $\kappa$ correspond to zeros (poles) of the $I_{ij}$.
This leads to \ac{RG} invariant relations between the physical parameters, which will be studied elsewhere.

\section{More details on Feruglio Model}
\label{sec:Feruglio_Model_1_modular_invariance_details}

The purpose of this appendix is to show that the observables in Feruglio's Model 1 (cf.\ \Cref{sec:Feruglio_Model_1}) are modular invariant, provided one transforms the \ac{VEV} of $\varphi_T$ appropriately. 
To see this, recall that the superpotential terms in this model are given by contractions between the flavon $\varphi_T$ and the triplet of modular forms. 
The invariance of superpotential terms  requires $A_{4}$ invariance and that the modular weights of fields and modular forms involved in  an operator to add up to zero. 
As for the former, the fact that $Y^{(2)}_\rep{3}$ transforms as a triplet means that 
\begin{equation}
  Y^{(2)}_\rep{3}(\gamma\,\tau)  
  = (c\tau+d)^2\,
  \rho_{\rep{3}}(\gamma)\,Y^{(2)}_\rep{3}\;.
\end{equation}
The fields, including the flavon $\varphi_T$, transform in such a way that the superpotential is invariant. 
The \ac{VEV} of the flavon $\varphi_T$ is given by $\Braket{\varphi_T} = (u,0,0)$.
Both this \ac{VEV} and the invariants $I_{ij}(\tau)$ from~\eqref{eq:I_ij_Feruglio_Model_1} are invariant under $T$ but not $S$ transformations. 
However, the transformations which change $I_{ij}(\tau)$ can be regarded as a basis change, and after undoing the basis change the invariants get mapped to their original form.

In slightly more detail, under an $S$ transformation
\begin{subequations}
\begin{align}
  Y^{(2)}_\rep{3}(\tau) =
    \begin{pmatrix}
    Y_1(\tau)  \\ Y_2(\tau)  \\Y_3(\tau) 
    \end{pmatrix}
  &\xmapsto{~S~}
  \tau^2 \rho(S)Y^{(2)}_\rep{3} (\tau) =\frac{\tau^2}{3}\,
    \begin{pmatrix}
    - Y_1 + 2 Y_2 + 2 Y_3 \\ 2 Y_1 - Y_2 + 2 Y_3 \\ 2 Y_1 + 2 Y_2 - Y_3
    \end{pmatrix}\;,\\
  M_e=u\,\diag(\alpha,\beta,\gamma)
  &\xmapsto{~S~}\frac{u}{3}\,
    \begin{pmatrix}
      -\alpha & 2\beta & 2\gamma\\
      2\alpha & -\beta & 2\gamma\\
      2\alpha & 2\beta & -\gamma
    \end{pmatrix}\;.\label{eq:Feruglio_Model_1_S_trafo_of_Y_e}
\end{align}
\end{subequations}
Under the $S$ transformation alone, $I_{ij}(\tau)$ from~\eqref{eq:I_ij_Feruglio_Model_1} are not invariant. 
However, once we diagonalize the charged lepton Yukawa couplings, which amounts to undoing~\eqref{eq:Feruglio_Model_1_S_trafo_of_Y_e},  $I_{ij}(\tau)$ get mapped back to their original form.
Of course, these findings are a simple consequence of two basic facts: 
(i) modular transformations of $Y(\tau)$ amount to transforming $Y(\tau)$ with an $A_4$ matrix and multiplying it by an automorphy factor, and (ii) invariants $I_{ij}$ are constructed in such a way that the automorphy factors cancel. 
Therefore, undoing the $A_4$ transformation returns $I_{ij}$ to their original form.

\bibliography{HolomorphicModular}
\bibliographystyle{utphys}

\begin{acronym}
\acro{2HDM}{two-Higgs doublet model}  
\acro{BSM}{beyond the standard model}
\acro{EFT}{effective field theory}
\acro{FCNC}{flavor changing neutral current}
\acro{IO}{inverted ordering}
\acro{MIHO}{modular invariant holomorphic observables}
\acro{MSSM}{minimal supersymmetric standard model}
\acro{NO}{normal ordering}
\acro{QFT}{quantum field theory}
\acro{RG}{renormalization group}
\acro{RGE}{renormalization group equation}
\acro{SM}{standard model}
\acro{SUSY}{supersymmetry}
\acro{UV}{ultraviolet}
\acro{VEV}{vacuum expectation value}
\end{acronym}
\end{document}